\documentclass[10pt]{article} 
\usepackage{ifpdf}
\usepackage{latexsym}
\usepackage{graphics}
\usepackage{epsfig}
\usepackage[dvips]{color}
\usepackage[mathscr]{eucal}
\usepackage{amscd}
\usepackage{amssymb}
\usepackage{amsmath}
\usepackage{amssymb}
\usepackage{amsthm}

\newtheorem{claim}{Claim}

\title{Evidence for a Phase Transition \\ in  2D Causal Set Quantum Gravity} 
\author{Sumati Surya,  \\ Raman Research Institute, Bangalore, India 
} 

\begin{document}
\maketitle

\begin{abstract} 
  We present evidence for a phase transition in a theory of 2D causal set quantum gravity which
  contains a dimensionless non-locality parameter $\epsilon \in (0,1]$.  The transition is
  between a continuum phase and a crystalline phase, characterised by a set of covariant
  observables. For a fixed size of the causal set the transition temperature $\beta_c^{-1}$
  decreases monotonotically with $\epsilon$. The line of phase transitions in the $\beta_c^2$ v/s
  $\epsilon$ plane asymptotes to the infinite temperature axis, suggesting that the continuum phase
  survives the analytic continuation.
\end{abstract}

Causal set theory(CST) is a discrete approach to quantum gravity which combines local Lorentz invariance 
with a fundamental discreteness \cite{blms}. The spacetime continuum is replaced by a locally finite
poset or causal set,  with the order relation $\prec$ being the analog of the
spacetime causal order. The continuum quantum gravity path integral is thus replaced by a 
discrete sum over causal sets  
\begin{equation} \label{part}
Z_{CST}=\sum_{C \in \Omega} \exp^{iS[C]/ \hbar},  
\end{equation}  
where $\Omega$ is a sample space of causal sets and $S[C]$ is an appropriately chosen action. The
combination of local Lorentz invariance with fundamental discreteness gives rise to a non-locality
in the continuum approximation, making the extraction of local geometric data highly
non-trivial. The recent construction of a discrete Einstein-Hilbert action, the Benincasa-Dowker
action for causal sets \cite{BD}, thus allows us for the first time to begin a serious study of the
causal set partition function $Z_{CST}$.

Apart from a choice of action, $Z_{CST}$ also depends crucially on the sample space $\Omega$. A
natural starting choice for $\Omega$ is the collection of countable causal sets; in classical
sequential growth models of causal sets, for example, $\Omega$ is further restricted to causal sets
that are past finite \cite{csg,observables}.  The collection of $N$-element causal sets $\Omega_N$
is known to be strongly dominated by the ``Kleitman-Rothschild''(KR) class of causal sets in the
large $N$ limit. These are of a fixed time extent with only three ``moments of time'', and admit no
continuum approximation \cite{kr}. This presents a potential ``entropy problem'' in CST.  In order
to be able to recover spacetime-like behaviour, therefore, the causal set action, or more generally
the choice of dynamics, should be able to counter this entropy. Indeed, classical sequential growth
dynamics is an example in which the KR entropy is made sub-dominant by the dynamics \cite{csg}.

In this work we consider a two dimensional theory of causal sets, defined by the 2-dimensional Benincasa-Dowker action $S_{2d}$ \cite{BD}, and an order theoretic dimensional restriction of $\Omega_N$ to $\Omega_{2D}$, the set of $N$-element ``2D orders'', defined as follows. Let $S=(1, \ldots, N)$ and $U=(u_1,u_2, \ldots u_N)$, $V=(v_1,v_2,\ldots v_N)$, with $u_i, v_i \in S $, $u_i\neq u_j$, $v_i \neq v_j$ for $i\neq j$. $U$ and $V$ are then {\sl total orders} with $\prec$ given by the natural ordering $<$ in $S$: for every pair $ i \neq j$ either $u_i < u_j$ or $u_j < u_i$, and similarly for $V$. An $N$-element 2D order is the {\sl intersection} $C=U\cap V$ of two total $N$-element orders $U$ and $V$, i.e., $e_i \prec e_j$ in $C$ iff $u_i < u_j$ {\it and} $v_i < v_j$. A useful example of an $N$-element 2D order is a set of $N$ events $\{e_1, \ldots, e_N\}$ in 2d Minkowski spacetime ordered by causality, and such that in light cone coordinates $e_i=(u_i, v_i)$, $u_i \neq u_j$ and $v_i \neq v_j$ for $i \neq j$.

The motivation for restricting the sample space to $\Omega_{2D}$ stems from the fact that the causal
set discretisation of a conformally flat, topologically trivial 2d spacetime is a 2D order
\cite{2dqg}. However, not all 2D orders can be approximated by continuum spacetimes. This means that
the choice of $\Omega_{2D}$ corresponds only to a restriction of poset dimension and not spacetime
dimension. Moreover, only in a very limited sense do these dimensions coincide: every 2D order $C$
admits an order preserving embedding $\Phi$ into a patch of 2d Minkowski spacetime $ ({}^2M,\eta)$, i.e.,
for every $e_i \prec e_j$ in $C$, $\Phi(e_i)$ causally precedes $\Phi(e_j)$ in $({}^2M,\eta)$. Such
an embedding though necessary, is not sufficient to ensure a continuum approximation for $C$.

A striking feature of 2D orders is that in the asymptotic limit $N\rightarrow \infty$, $\Omega_{2D}$
is dominated by ``random'' 2D orders, namely those approximated by Minkowski spacetime
\cite{2dqg,elsauer,winkler}. A random 2D order is the intersection of two total orders $U=(u_1,
\ldots, u_N)$ and $V=(v_1, \ldots v_N)$ which are chosen randomly and independently from $S$.  Hence
unlike the unrestricted sample space, spacetime like causal sets dominate the uniform measure on
$\Omega_{2D}$. It is thus of obvious interest to study the effect of $S_{2d}$ on this entropic
feature of $\Omega_{2D}$.

\begin{figure}[ht] 
\centering \resizebox{2.5in}{!}
{\includegraphics{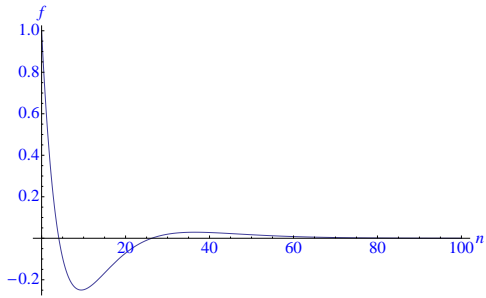}}
\vspace{0.5cm}
\caption{{\small The function $f(n,\epsilon)$ for $\epsilon=0.12$.}}\label{fne}
\end{figure}

While there is no natural Planck scale in 2D gravity, in CST one requires a volume cut-off
$V_p=l_p^2$ in order to realise the continuum approximation, and this plays the role of a
fundamental scale.  In addition, the 2d Benincasa-Dowker action $S_{2d}[C]$ for a causal set $C$
includes a ``non-locality'' scale $l_k>l_p$ required to suppress large fluctuations about the mean
in the continuum approximation \cite{BD,dalem}.  $S_{2d}[C]$ can be expressed in terms of the
abundances $N_n$ of the ``intervals'' $I[i,j] \equiv \{ k| i\prec k \prec j \}$ of fixed cardinality
$n$ in $C$
\begin{equation} \label{action} 
\frac{1}{\hbar} S_{2d}[C,\epsilon]= 4 \epsilon (N - 2 \epsilon \sum_{n=0}^{N-2} N_n f(n,\epsilon)), 
\end{equation}
where $\epsilon=l_p^2/l_k^2$ and 
\begin{equation} 
f(n,\epsilon)=(1-\epsilon)^n\biggl(1-\frac{2\epsilon
  n}{(1-\epsilon)}+\frac{\epsilon^2n(n-1)}{2 (1-\epsilon)^2}\biggr).
\end{equation} 
Figure \ref{fne} shows the typical behaviour of $f(n,\epsilon)$. 

For $n > 1/\epsilon$, $f(n,\epsilon) \sim 0$, and thus, effectively, the action for a fixed
$\epsilon$ includes the abundance of intervals only upto size $1/\epsilon$. To avoid infrared errors
in counting such intervals for finite $N$, $\epsilon$ should in addition be bounded below by $N
\epsilon >1$.

The goal of writing down a partition function for quantum gravity is to eventually construct and
calculate expectation values of covariant observables. The analog of covariance in causal set theory
is label independence, and hence we seek to construct label invariant observables. We have already
encountered an example of such an observable, namely, the action $S_{2d}/\hbar$.  Another important
covariant observable is the Myrheim-Myer dimension \cite{mm} for a causal set, which in the
continuum approximation reproduces the Minkowski spacetime dimension. We will construct several such covariant
observables in what follows and use them to determine the existence of a continuum approximation.

However, as we have defined it, the sample space $\Omega_{2d}$ is itself the collection of {\it
  labelled} 2D orders. While every 2D order admits $N!$ worth of relabellings, each of these
relabellings does not produce a distinct labelled 2D order. If $C$ is a labelled causal set
containing a pair of elements $e_i$ and $e_j$ with the same past and future sets, the relabelling $
i \leftrightarrow j $ is an {\it automorphism} $\Pi_{i\leftrightarrow j}: C \rightarrow C$. An
extreme example is the antichain or completely unordered set -- {\it every} relabelling produces the
same labelled causal set. Since the action itself is label invariant, the partition function over
unlabelled causal sets has an additional weight of $N!/|Aut(C)|$ for each unlabelled 2D order $C$,
where $|Aut(C)|$ is the cardinality of the Automorphism group of a labelled counterpart. While it is
possible to view this additional measure as a ``quantisation ambiguity'', it should be
stressed that the observables that we construct are nevertheless {\it strictly} covariant.

An approach to evaluating the expectation values of covariant observables is to Euclideanise the
partition function, in the process converting the quantum system into a thermodynamic one.
Replacing the set of Lorentzian metrics wholesale for Euclidean ones, however, makes little of the
importance of causal structure, and moreover, can lead to highly fractal, non-manifold like
behaviour \cite{dt}. Instead, as suggested in \cite{euclidean} one can transition to a thermodynamic
partition function by introducing a new parameter $\beta$ into the action and taking $\beta
\rightarrow i \beta$.  This leaves the sample space unchanged and hence there is no ambiguity in the
interpretation of the covariant thermodynamic results. A similar ``parameter-based'' analytic
continuation employed in the causal dynamical triangulation approach, on the other hand, does lead
to a Euclideanisation of the sample space itself \cite{cdt}.

For concreteness, we adopt the following prescription. Namely, we replace the complex weights
$\exp^{i S_{2d}/\hbar}$ with $\exp^{i \beta S_{2d}/\hbar}$, and 
obtain the thermodynamic partition function by taking $\beta \rightarrow i \beta$:  
\begin{equation} 
Z_N=\sum_{C\in \Omega_{2D}} \exp^{-\beta S_{2d}[C]/\hbar}. 
\end{equation} 
The $\beta \rightarrow 0$ limit is the uniform distribution, dominated by the 2D random orders,
approximated by an interval in Minkowski spacetime for finite $N$. As $\beta \rightarrow \infty$,
since the action is not positive definite, causal sets with the largest negative values of the
action should dominate,  modulo entropic effects. Thus, one expects a cross-over at finite $\beta$.

In this work, we use Markov Chain Monte Carlo(MCMC) methods to study this partition function. The
results we present are the first in a larger effort to study causal set quantum dynamics using MCMC
techniques \cite{ongoing}. The restriction to 2 dimensions leads to a substantial simplification
which translates into rapid mixing or thermalisation of the Markov Chain. Our simulations are
carried out for relatively small causal sets ($N=50$) but this is sufficient to show emergent
continuum behaviour.  One of the main observations of our present work is the existence of a phase
transition at finite $\beta$, rather than the cross-over suggested above. The transition is from a
continuum-like phase to a crysalline non-continuum phase, and is well characterised by the change in
the expectation values of a set of covariant observables.

We first briefly review the basics of causal set theory, and refer the reader to
\cite{cstreferences} for more detail. A causal set is a locally finite partially ordered set
(poset), i.e., a countable collection of elements with an order relation $\prec$ which is (i)
transitive ($x \prec y\, , y \prec z \, \Rightarrow x \prec z$), (ii) {\sl irreflexive}, ($x
\not\prec x$) and (iii) locally finite, i.e., if $Past(x) \equiv \{w \in C| w \prec x \}$ and
$Fut(x)\equiv \{w \in C| w \succ x \} $ then the cardinality of the set $Past(x) \cap Fut(y)$ is
finite.  We say that two elements are {\sl {linked}} if $x \prec y $ and there is no $z$ such that
$x \prec z \prec y$. Local finiteness implements the physical requirement of a fundamental spacetime
discreteness: a finite spacetime volume contains only a finite number of ``spacetime atoms''.  A
continuum spacetime $(M,g) $ is said to be an {\it approximation} to an underlying discrete causal
set $C$ for a spacetime volume ``cut-off'' $V_c$, if there exists an embedding $\Phi:C\rightarrow
(M,g)$ which is (i) order preserving: for $x,y \in C$, $x \prec y \Leftrightarrow \Phi(x) \prec_M
\Phi(y)$, and (ii) $\Phi(C)$ is a Poisson distribution of events in $(M,g)$:
\begin{equation} \label{poisson} 
P_V(n)= \frac{(\rho V)^n} {n!} \exp^{-\frac{V}{V_c}}
\end{equation}   
is the probability of finding $n$ elements of $\Phi(C)$ in a spacetime volume $V$. Conversely, one
can generate a causal set $C$ from $(M,g)$ via a Poisson sprinkling, with $\prec$ induced by
$\prec_M$ for the sprinkled points. A causal set generated this way is therefore not a regular
lattice but a ``random lattice''. Starting with a sample space of finite element causal sets
$\Omega_N$, the thermodynamic limit $N \rightarrow \infty$ does not correspond to the continuum
limit of the theory, but rather its infrared limit. This means that a finite cardinality causal set
can be well approximated by a region of continuum spacetime without taking the $N \rightarrow
\infty$ limit.

We construct the MCMC via the {\it exchange move} on a 2D order $C=U \cap V$ defined as
follows. First pick $U$ or $V$ at random.  Wlog, let this be $U$.  Next, pick a pair $(u_i,u_j)$ of
elements of $U$ at random and perform the exchange $u_i \leftrightarrow u_j$, while leaving $V$
unchanged.  The new 2D order then has the new elements $e_i'=(u_i',v_i')=(u_j,v_i)$ and
$e_j'=(u_j',v_j')=(u_i,v_j)$, while all other elements remain the same. It is clear that the move is
reversible.  
\begin{claim} 
The exchange move on labelled 2D orders has no fixed points in the space $\Omega_{2D}$ of labelled  2D orders. 
\end{claim} 
\noindent {\it Proof:} Wlog, let $u_i \leftrightarrow u_j$ for some $i \neq j$. There are three
cases to examine: (a) If $e_i \prec e_j$, then $u_i < u_j$ and $v_i <v_j$. Under the exchange
$u_i'>u_j'$ while $v_i'<v_j'$ which means that $e_i'$ and $e_j'$ are now unrelated. (b) Similarly if
$e_i \succ e_j$, then after the exchange the two elements are unrelated. (c) If they are unrelated
to start with then either (i) $u_i<u_j$ and $v_i>v_j$, in which case after the exchange $e_j'<e_i'$
or (ii) $u_i>u_j$ and $v_i<v_j$, in which case after the exchange $e_j'> e_i'$. In all three cases,
therefore the exhange move makes a non-trivial modification to the relationship between the $i$th
and $j$th elements, thus giving rise to a new, distinct labelled causal set.  \hfill $\qed$ 

We employ a Metropolis-Hastings algorithm, accepting a move if the difference in the action $\Delta
S_{2d}$ is negative and rejecting a move only if $\exp^{- \beta \Delta S_{2d}/\hbar} < r$, where $r$
is a random number in $[0,1)$. Since each move depends on a pair of elements, we define a sweep to
be $N(N-1)/2$ moves, with the following observables recorded every sweep. (i) The ordering fraction
$\chi = 2r/N(N-1) $ where $r$ is the number of relations in the causal set with $N(N-1)/2$ the
maximum number of relations possible; $\chi$ is therefore analogous to the filling fraction in an
Ising model. In 2 spacetime dimensions $\chi$ is also the inverse of the Myrheim-Myer
dimension. Thus, for flat spacetime, we should expect $\langle \chi \rangle \sim 1/2$. (ii) The
action $S_{2d}[C]/\hbar$: for flat space it was shown in \cite{GB} that $ \langle S_{2d}/\hbar
\rangle \sim 4 $ for $\epsilon=1$ and simulations show that this is also true for $\epsilon \neq 1$.
(iii) Time Asymmetry: given that the action is itself time-reversal invariant, it is useful to check
whether the dynamics preserves this property. A rough measure of time asymmetry $t_{as}$ is the
difference in the number of maximal and minimal elements in the causal set; in flat spacetime
$\langle t_{as} \rangle \sim 0$. (iv) The height $h$ of the causal set or the length of the longest
chain. This corresponds in the continuum approximation to the maximum proper time in flat spacetime
\cite{BG}.  (v) The abundances $N_n$ of $n$-element intervals for all $n \in [0, N-2]$. In flat
spacetime $N_n$ has a specific monotonic fall off with $n$ and this provides a useful comparison.

Our focus in this work is to examine the behaviour of these observables as function of the
inverse temperature $\beta$  as one varies the non-locality parameter $\epsilon$. We present results for
$N=50$, with a range of values for $\epsilon$ between $0.1$ and $1$. We perform the simulations for
$10,000$ sweeps which translates to $12.5$ million attempted moves.

\begin{figure}[ht] 
\centering \resizebox{2in}{!}{\includegraphics{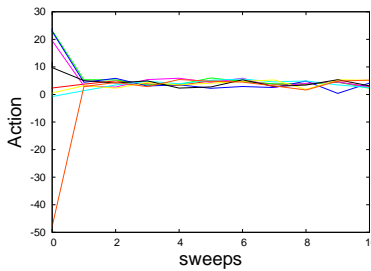}}
\vspace{0.5cm}
\caption{{\small The exchange move gives rise to a rapid thermalisation starting from the eight different different
    starting 2D orders. Here we show a plot of action v/s number of sweeps for $\epsilon=0.12$
    and $\beta=0.1$. Thermalisation takes place by the 2nd sweep. }}\label{ergodicity}
\end{figure}

We test for thermalisation starting from several different types of initial 2D orders. Our current
algorithm includes the following 8 initial types of causal sets: (i) a randomly labelled chain or totally ordered
set (ii) the antichain (iii) a random 2D order, which is approximated by the Minkowski interval
(iv-v) two 2D random orders of size $N/3$ each with an intervening $N/3$ element chain or
antichain, (vi) two 2D random orders of size N/16 with an intervening antichain of size 7N/8 (vii)
two $N/2$ element random orders stacked on top of each other and (viii) a crystalline causal set of
the kind we will describe shortly. We find that there is rapid mixing starting from these 
varied configurations, an example of which is shown in Figure \ref{ergodicity}.
The expectation values $\langle \; \mathcal O \;\rangle$ are calculated from a sampling per
autocorrelation time $\tau_{\mathcal O}$. As expected, $\tau_{\mathcal O}$ increases with $\beta$ so
that the number of independent samples rapidly decreases for larger $\beta$. 

\begin{figure}[ht]
\centering \resizebox{2.25in}{!}{\includegraphics{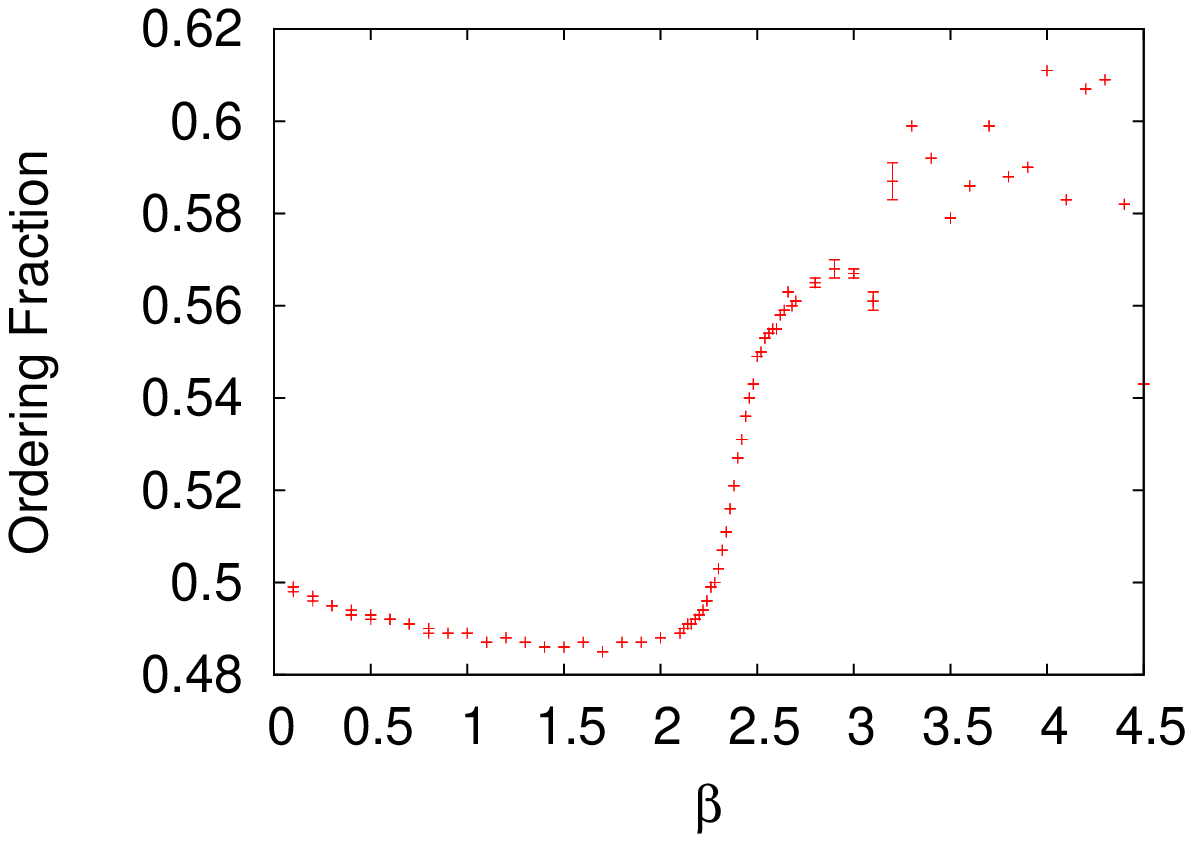}}
\centering \resizebox{2.25in}{!}{\includegraphics{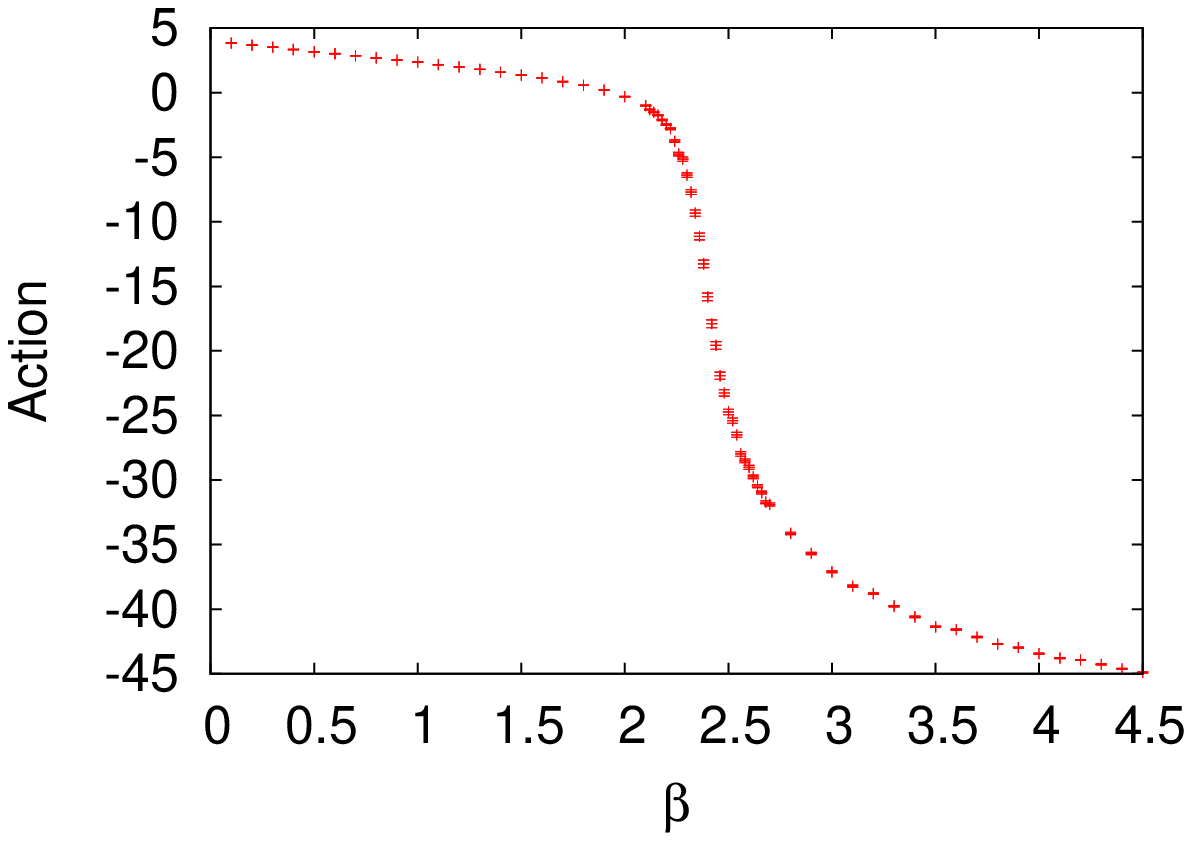}}
\centering \resizebox{2.25in}{!}{\includegraphics{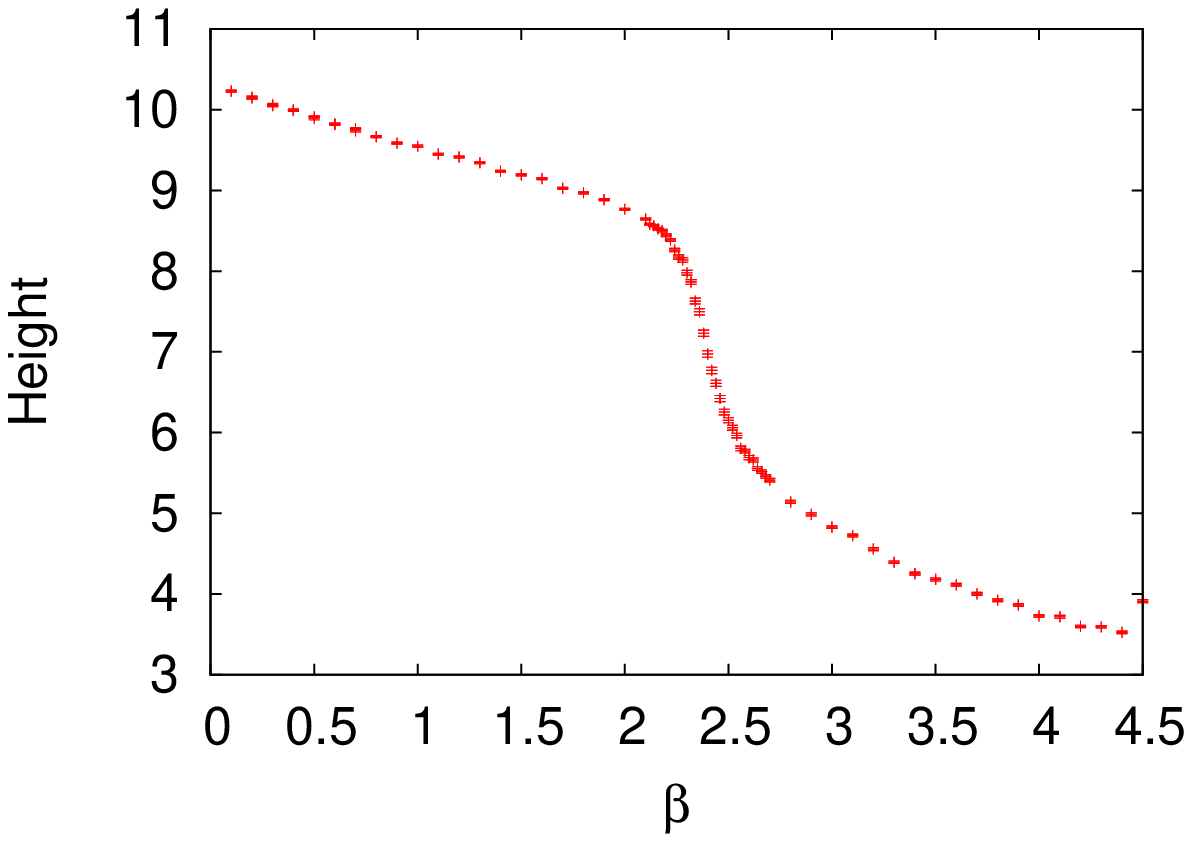}}
\centering \resizebox{2.25in}{!}{\includegraphics{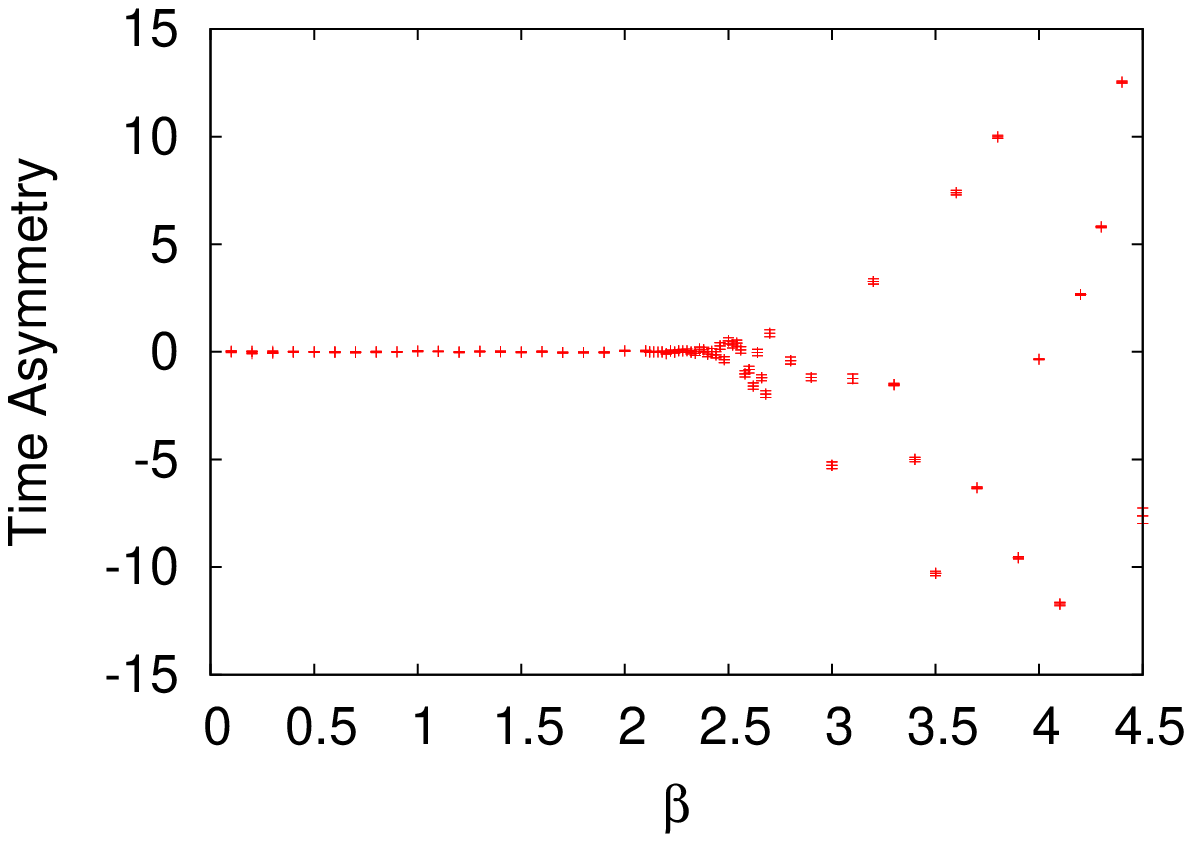}}
\centering \resizebox{2.25in}{!}{\includegraphics{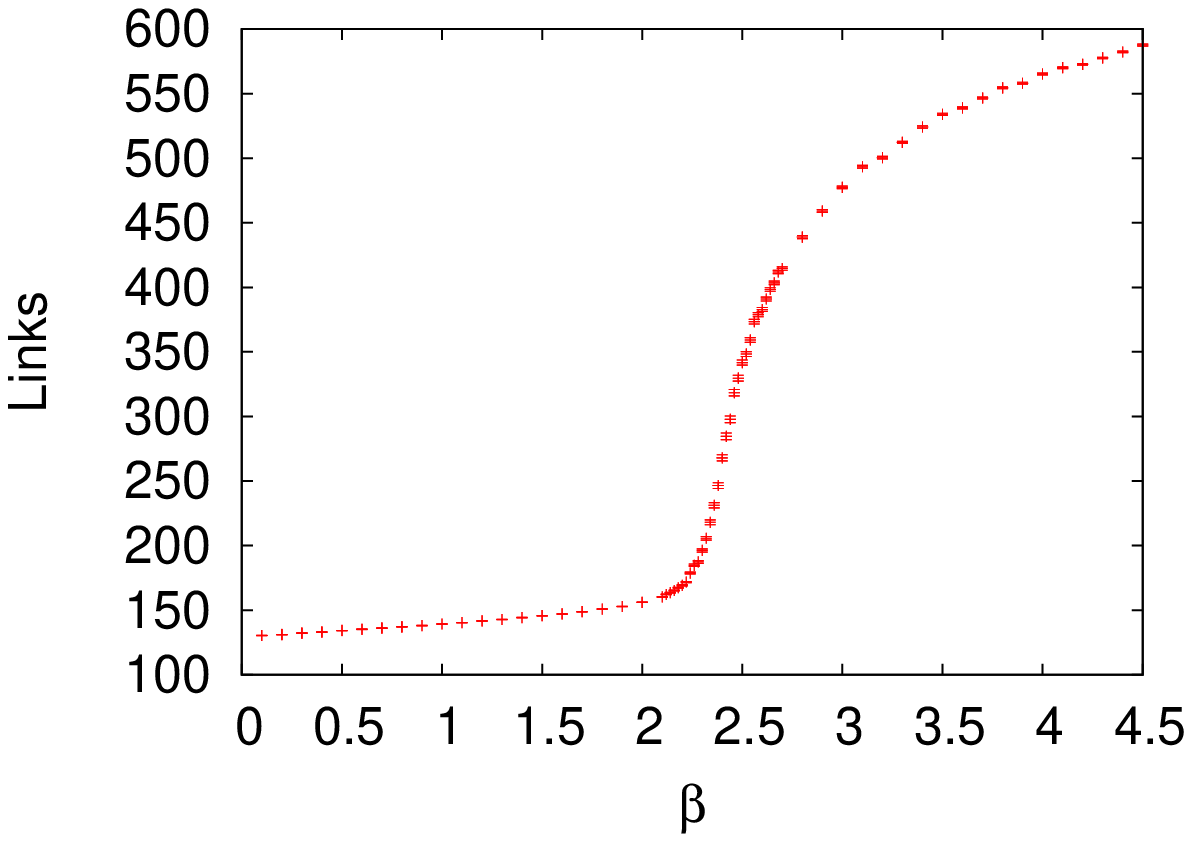}}
\caption{{\small Observables v/s $\beta$ for $N=50, \epsilon=0.12$. }}
\label{observables}
\end{figure}

For fixed $\epsilon$ we find a rapid change in the expectation values of the observables around a
fixed $\beta=\beta_c$ suggesting a phase transition as shown in Figure (\ref{observables}). The
exact value of $\beta_c$ can be determined by looking at the behaviour of the autocorrelation times
which peak at the same $\beta_c$ for all observables.  For the example shown in Figure
\ref{observables} it is clear that roughly, $2.2 < \beta_c < 2.5$.  We have included the error bars
in the plots, but these are typically very small.

\begin{figure}[ht] 
\resizebox{1.75in}{!}{\includegraphics{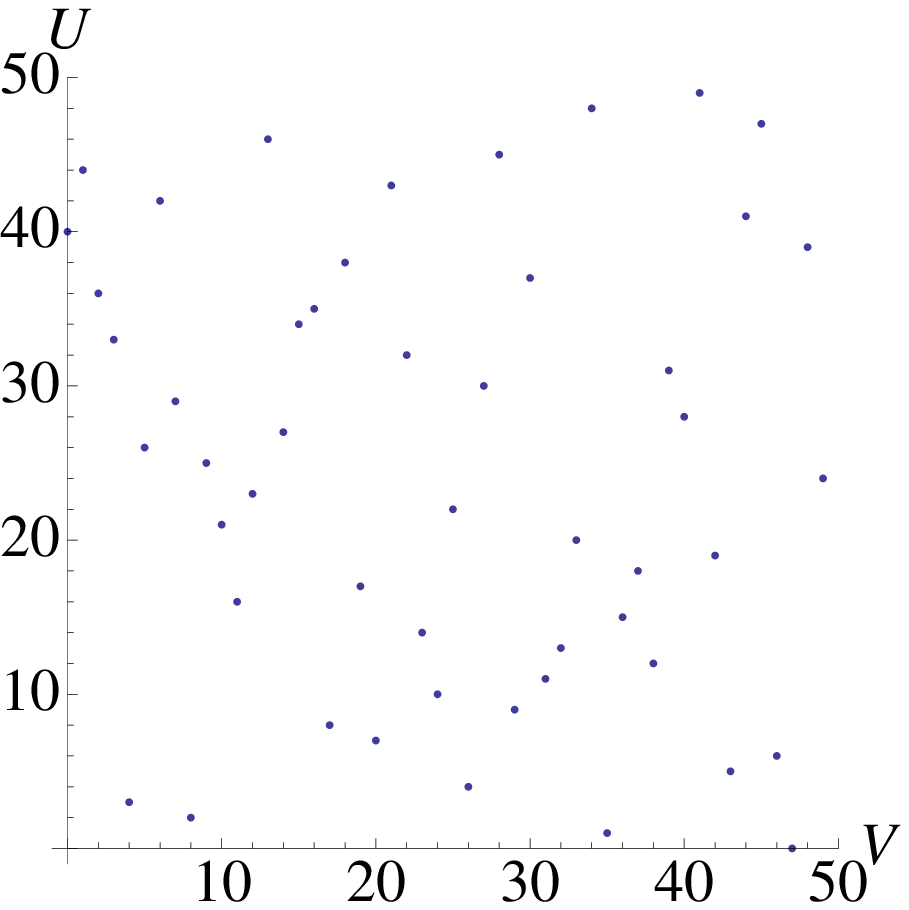}}\hspace{3cm}
\resizebox{1.75in}{!}{\includegraphics{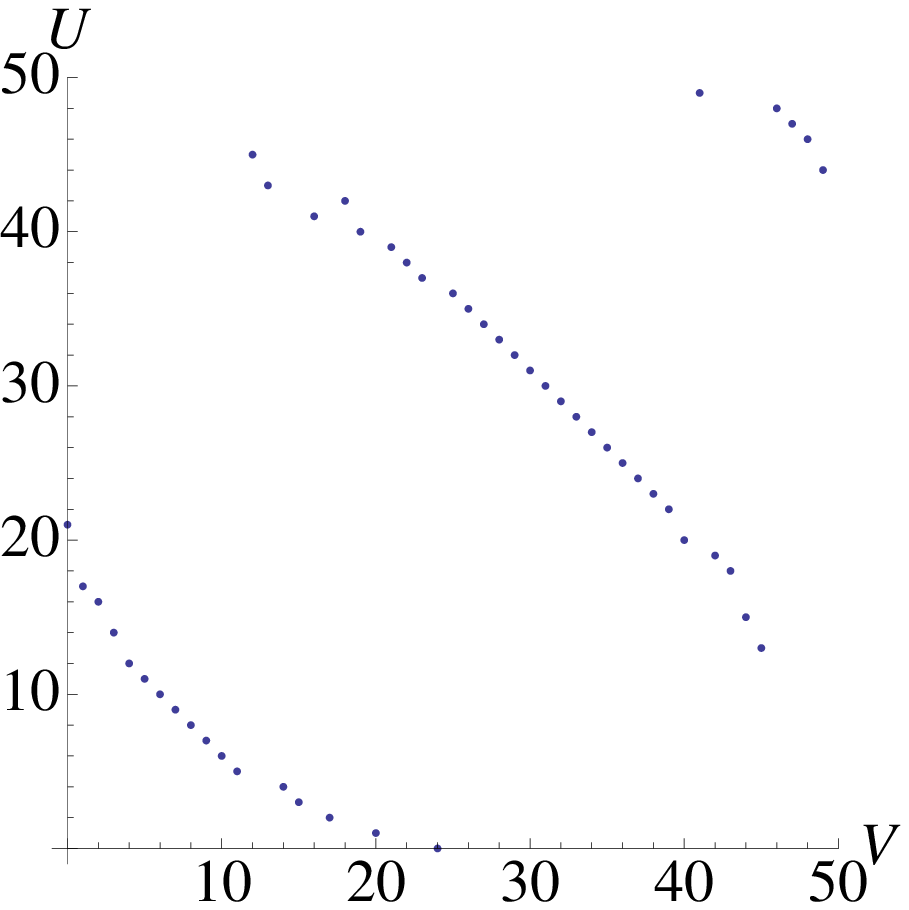}}
\caption{{\small On the left is a causal set in the Continuum Phase and on  the right one in the Crystalline Phase. $(U,V)$ are light cone coordinates that have been rotated clockwise by $\pi/4$. }}\label{phases}
\end{figure}

That there is an actual differentiation into two phases becomes explicit on examining the 2D orders
themselves. We record the configurations in the Markov Chain every 100 sweeps.  For $\beta<
\beta_c$, we find evidence for a ``continuum'' phase (Phase I), and for $\beta > \beta_c$, a
``crystalline'' phase (Phase II). We show examples of configurations in these two phases in Figure
\ref{phases}, where the light cone coordinates have been turned clockwise by $\pi/4$ for ease of
plotting.

\begin{figure}[ht] 
\centering \resizebox{3in}{!}{\includegraphics{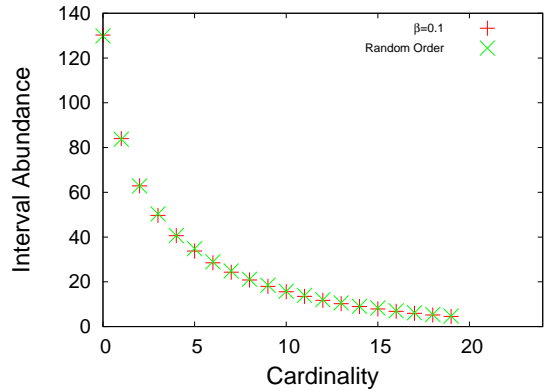}}     
\caption{{\small The distribution  of $N_n$ in $C_I$. This is compared with the distribution for a 
    2D Random order which is a discretisation of Minkowski spacetime. }}\label{abundance}
\end{figure}

A quick look at a typical causal set $C_I$ from Phase I shows a causal set that resembles a random
discretisation of a spacetime. Indeed, the expectation values for a fixed $\beta< \beta_c$
corroborate this. For example for $\epsilon=0.12$ and $\beta=0.1$, (a) ordering fraction:
$<\chi>=0.499$ with an error less than $10^{-3}$ , which means that the Myrheim-Myer dimension
$<d_{MM}> \sim 2.004$, (b) height: $<h>=10.232 \pm 0.014$, which should be compared to the height of
a $V=50 l_p^2$ Minkowski interval which is $10 l_p$, (c) time asymmetry: $<t_{as}>= 0.027 \pm 0.024$
and the (d) Action: $<S>/\hbar= 3.846 \pm 0.013$ which should be compared to the residual or
boundary value of $4$ for Minkowski spacetime \cite{GB}. In addition, if we plot the abundance $N_n$
of the intervals of cardinality $n$ in $C_I$ and contrast it that for a random 2D order, we find
that these match very closely as shown in Figure \ref{abundance}. As $\beta$ nears the transition,
the expectation values of these observables gradually change. However, the typical causal set
continues to retain the features of a random lattice.
Thus, looking at the explicit values of the observables, it is clear that Phase I corresponds to a
continuum phase. 

\begin{figure}[ht]
\centering \resizebox{3in}{!}   {\includegraphics{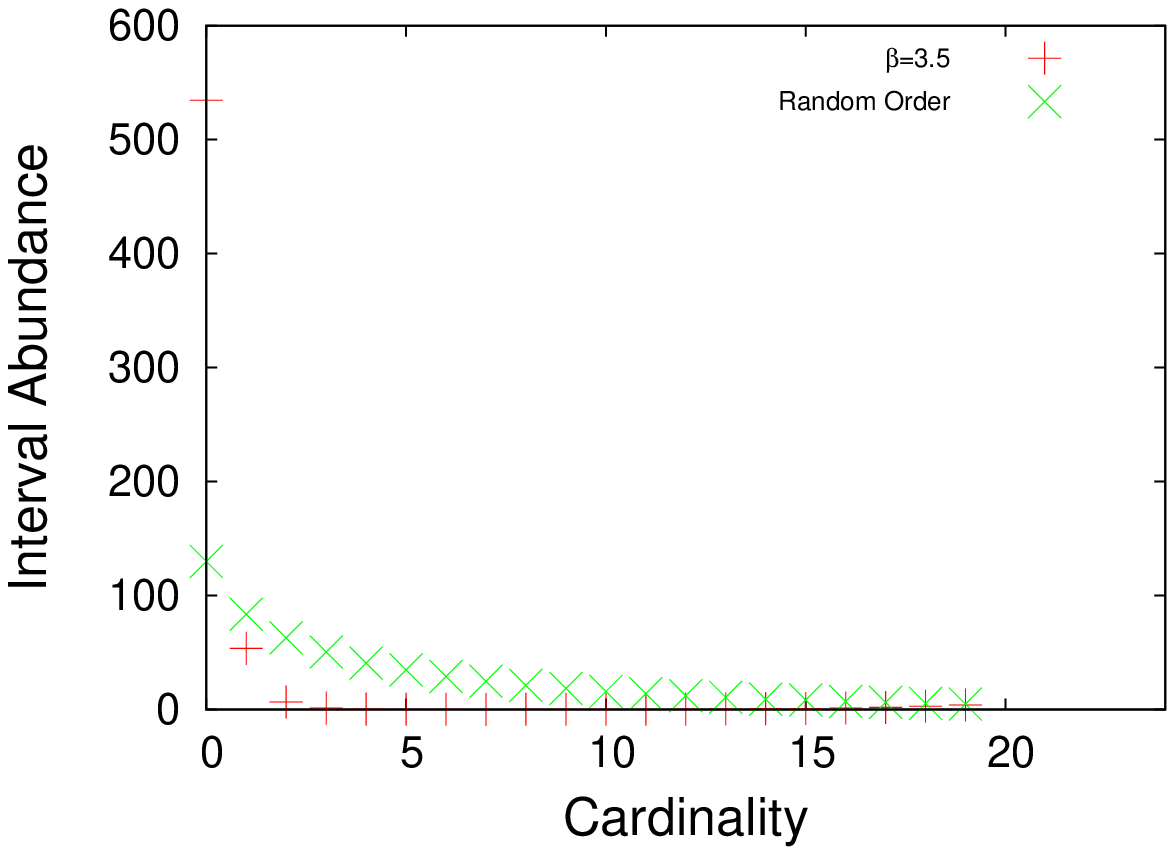}} 
\centering \resizebox{3in}{!}   {\includegraphics{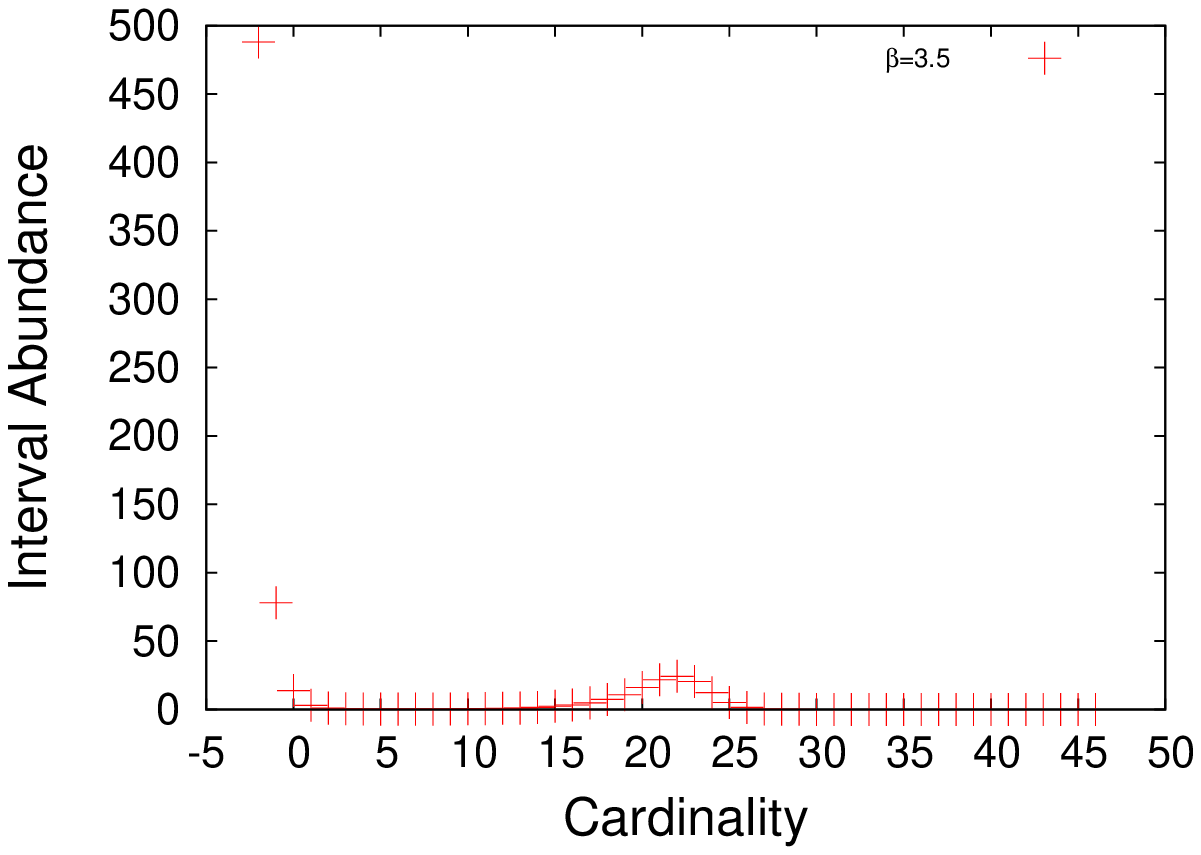}}
\caption{{\small The distribution of $N_n$ in $C_{II}$. On the left it is compared to a 2D Random order for
    $n<20$. On the right it is shown for all $n$ and can be seen to roughly follow the positive part of
    $f(n,\epsilon)$.}}\label{crysabundance}
\end{figure}

A typical causal set $C_{II}$ in Phase II on the other hand, has a most unexpected character as
shown in Figure \ref{phases}. It shares some superficial features of a KR causal
set, being of limited time extent, but does not strictly belong to this class. It has a regularity,
or crystalline nature, suggesting that it does not have a continuum approximation. The values of
most observables differ considerably from those in Phase I.  For example, for $\epsilon=0.12$,
$\beta=3.5$ (a) Ordering Fraction: $<\chi>=0.579$ which gives a fractal dimension for the causal set
of $<d_{MM}> \sim 1.727$, (b) Height: $<h>=4.180\pm 0.018$ (c) Time Asymmetry: $<t_{as}>= -10.292
\pm 0.102$, which is a large deviation from the expected value of zero. This may be the result of a
spontaneous breaking of the time symmetry, similar to that in the low temperature phase of the Ising
model, but may also be the result of poorer statistics at larger $\beta$. (d) Action: $<S>/\hbar=
-41.367 \pm 0.054$, i.e., the action tries to take on the lowest possible (negative) value. In
contrast to Phase I, the abundances of intervals as shown in Figure \ref{crysabundance} is very
different from that of Minkowski spacetime. In particular, $N_n$ align themselves with the positive
part of $f(n,\epsilon)$ and vanish for those $n$ for which $f(n,\epsilon) <0$, thus minimising the
action.

The existence of these distinct phases thus strongly suggest a phase transition. It is tempting to
draw the obvious analogy with the Ising model: at high temperatures one has the disordered or random
Phase I, while at low temperatures, there is the higly ordered crystalline Phase II which exhibits a
spontaneous breaking of symmetry.  Although it is difficult at this stage of our work to assess the
order of this transition, there are hints that it may be of second order. To begin with, the
autocorrelation time peaks at the phase transition as do the fluctuations in the observables. The
size of the fluctuations gives us a rough estimate of the temperature of the phase transition, and
this coincides with the peak in the autocorrelation time. Moreover, the transitions shown in Figure
\ref{observables} appear to be smooth. A more conclusive assessment would require a detailed
understanding of the dependence of these results on  the cardinality $N$. 

It is important to stress that while the nature of this phase transition is definitely of interest,
it does not play a crucial role in determining continuum behaviour. In other lattice-based
approaches in which discretisation is used only a calculational tool, the appearance of a second
order phase transition signals the fact that the continuum limit of the theory exists. However, in a
fundamentally discrete theory like causal set theory, it is only the continuum approximation we
seek, and as we have just seen, this does not depend on the existence and the nature of a phase
transition.

\begin{figure}[ht]
\centering \resizebox{3in}{!}{\includegraphics{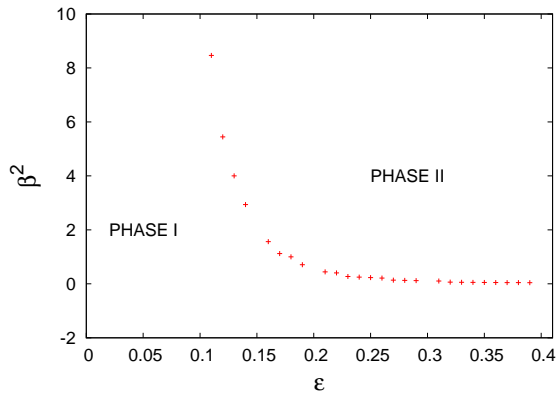}}
\vspace{0.5cm}
\caption{{\small This is a plot of the locus of critical points $\beta_c^2$ as a function of
    $\epsilon$. For $\beta^2>\beta_c^2$ the system is in the crystalline phase and for
    $\beta^2<\beta_c^2$ it is in the continuum phase.}}\label{criticalline}
\end{figure}

More important to our discussion, then is the question of what this thermodynamic calculation can
mean for quantum gravity. How much of the above discussion, if any, survives the analytic
continuation? As $\epsilon$ varies in $(0,1]$, our simulations show that the phase transition
survives, but the critical temperature increases monotonically with $\epsilon$.  Using the maximal
size of the fluctuations to estimate the critical temperature, we plot $\beta_c^2$ as a function of
$\epsilon$. The negative $\beta^2$ axis corresponds to the region of interest, i.e., the quantum
regime, while the positive $\beta^2$ axis corresponds to the thermodynamic regime to which our above
calculations belong. As shown in Figure \ref{criticalline} the line of phase transitions which
separates Phase I from Phase II asymptotes to the $\beta^2=0$ axis which itself belongs to the
continuum Phase I.  This strongly suggests that the non-continuum crystalline Phase II is confined
to the $\beta^2>0$ region and that the continuum phase survives the analytic continuation into the
$\beta^2<0$ region. The phase diagram also suggests that smaller $\epsilon$ is in a sense
``favoured'' by Phase I, which is consistent with having smaller fluctuations in the action and
hence a more reliable continuum approximation. A similar type of analysis has been used recently in
studying the phase structure of QCD \cite{forcrand}.

A more careful study of the phase diagram is clearly in order. For example, does the gradual change in the observables in Phase I as one moves towards $\beta_c$ signal a significant shift away from Minkowski spacetime?  In a recent paper \cite{ajjl} the phase structure of the 4d causal dynamical triangulation model of quantum gravity was studied. The phase diagram of this theory includes three phases, two of which, Phase C and Phase B, seem to be at least superficially analogous to our Phases I and II respectively.  The phase transition B-C is argued in \cite{ajjl} to be second order and it would be interesting to explore whether these two theories lie in the same universality class.

Since a more ambitious goal is to work with the unrestricted sample space $\Omega$, and an action
with at least dimension $4$ \cite{ongoing}, it is not immediate that 2d CST can teach us
straightforward lessons. Nevertheless, our analysis opens a new window into causal sets, and with it
a host of questions that can finally begin to be addressed. One of the more interesting of these is
whether there is a renormalisation group type analysis with stable fixed points for
$\epsilon<1$. This would suggest that the non-locality scale is not a free parameter, but can be
determined from the quantum dynamics.

\vskip 0.5cm
\noindent {\bf Acknowledgements:} I would like to thank Rafael Sorkin, David Rideout, Fay Dowker and
Alexei Kurkela for discussions. I am also grateful for the hospitality of the High Energy Theory
Group at McGill University where most of this work was carried out.  This
research was supported in part by an NSERC Discovery grant to the McGill University High Energy
Theory Group and an ONR grant to McGill University. The numerical simulations were carried out on a
high performance cluster at the Raman Research Institute.

\end{document}